\documentclass[superscriptaddress, amsmath,amssymb, aps, prb, twocolumn]{revtex4}
\usepackage[
                 colorlinks = true,  
                 linkcolor = blue,
                 urlcolor  = blue,
                 citecolor = blue
                 ]{hyperref} 
\usepackage{enumerate}
\usepackage{tikz}
\usetikzlibrary{patterns}
\usepackage[caption=false,singlelinecheck=off]{subfig}
\usepackage{slashed}
\usepackage{braket}
\usepackage{subfloat}
\usepackage{bbold}
\usepackage{booktabs}
\newcommand{\n}{\nu}
\usepackage{blindtext}
\usepackage{tikz}\usetikzlibrary{plotmarks}
\usepackage{multirow}
\usepackage{bbm}
\usepackage{graphicx}
\usepackage{dcolumn}
\usepackage{bm}
\usepackage[mathlines]{lineno}
\usepackage[normalem]{ulem}

\begin{document}
\title{Non-local order parameters for states with topological electromagnetic response}
\author{Thomas Klein Kvorning}
\affiliation{Department of Physics, University of California, Berkeley, CA 94720, USA}
\author{Christian Sp\r{a}nsl\"{a}tt}
\affiliation{Department of Physics, Stockholm University, SE-106 91 Stockholm, Sweden}
\affiliation{Institut f\"{u}r Nanotechnologie, Karlsruhe Institute of Technology, 76021 Karlsruhe, Germany}
\affiliation{Institut f\"{u}r Theorie der Kondensierte Materie, Karlsruhe Institute of Technology, 76128 Karlsruhe, Germany}
\author{AtMa P.O. Chan}
\affiliation{Department of Physics and Institute for Condensed Matter Theory, University of Illinois at Urbana-Champaign, Illinois 61801, USA}
\author{Shinsei Ryu}
\affiliation{James Franck Institute and Kadanoff Center for Theoretical
Physics, University of Chicago, Illinois 60637, USA}

\date{\today}
\begin{abstract}
Chern insulators are states of matter characterized by a quantized Hall
conductance, gapless edge modes but also a singular response to monopole
configurations of an external electromagnetic field.
In this paper, we describe the nature of such a singular response and show how it can be used to define a class of operators acting as non-local order parameters. These operators characterize the Chern-insulator states in the following way: for a given state, there exists a corresponding operator which has an algebraically decaying two-point function in that particular state, while it decays exponentially in all other states. The behaviour of the order parameter is defined only in terms of the electromagnetic response, and not from any microscopic properties, and we therefore claim to have found a generic order parameter for the Chern insulating states. We support this claim by numerically evaluating the order parameters for different insulating states. We also show how our construction can be generalized to other states with topological electromagnetic response, and use the states with a quantized magnetoelectric effect in three dimensions as an example. Besides providing novel insights into topological states of matter, our construction can be exploited to efficiently diagnose such states numerically.
\end{abstract}
\maketitle
\section{Introduction}
\label{sec:Introduction}
Topological states of matter can, by definition, not be assigned any symmetry breaking local order parameter. Instead, their characteristics are captured by global properties, i.e. topological invariants~\cite{Chiu2016}, or equivalently, universal properties of the edge-physics~\cite{Callan1985,Stone1991,Ryu2012}.

One may however ask whether there exists some non-local transformation to a dual theory, where the topological order turns into a local order, i.e. an order characterized by correlations functions of local operators. Such a formulation could be beneficial by providing novel ways to view known phenomena, but might also be a method to discover new ones that might be difficult to find with other means. 

The main question we address in this work is whether such non-local transformations can be found just by knowing the low energy response to external perturbations. We consider states characterized by a topological response to external fields coupling to a conserved $U(1)$ (e.g. electromagnetic) current. For concreteness, our primary focus are the Chern insulators~\cite{Haldane1988}, which are spatially two-dimensional ($2+1D$ space-time dimensions) states of matter characterized by their quantized Hall conductance. However, our approach is quite general, and should be possible to generalize to many other states in the symmetry protected topological periodic table~\cite{Altland1997,Qi2008,Ludwig2008,Kitaev2009,Ryu2010}. As an example of such a generalization,  we consider states with a quantized magnetoelectric effect in $3+1D$~\cite{Fu2007,Moore2007,Roy2009,Ryu2010,Hosur2010,Chan2016,essin2009}.

The Chern insulators are topologically equivalent to the integer quantum
Hall (IQH) states~\cite{Klitzing1980,Laughlin1981} (i.e. an integer number
of filled Landau levels) and therefore share the same $U(1)$ response.
For the IQH states, one can explicitly show that
they can be understood as states of condensed composite bosons. These
bosons are defined by a singular flux attachment to the original electrons which changes the exchange statistics from fermionic to bosonic. Such a ``transmutation'' can be understood as follows: in $2+1D$, each electron of the system can be viewed as a hard core boson attached to a fictitious flux quantum. This flux quantum can be taken to point in the opposite direction of the real magnetic field and in a filled Landau level, the fictitious field will on average cancel against the real magnetic field. The system can thus, on large length-scales, be viewed as consisting of a condensation of free ``composite bosons''.

Such a Chern-Simons-Bose-Einstein condensate, as described in Refs.~\onlinecite{Zhang1989,Kivelson1992}, is characterized by algebraic decaying two-point boson correlators. More precisely, the second quantized composite boson operator $\phi$ exhibits the following property
\begin{equation}
\Braket{\phi^{\dagger}(\mathbf{x}^{\prime})\phi(\mathbf{x})}\propto|\mathbf{x}-\mathbf{x}^{\prime}|^{-\alpha},
\end{equation}
where $\alpha$ is a constant shown below to be directly related to the long-range $U(1)$ response. The filled Landau level, can therefore be viewed as a long range ordered bosonic condensate as demonstrated by the existence of the algebraically decaying correlation functions~\cite{Girvin1987}. In this work, we view such an operator as a non-local order-parameter, or just order-parameter, keeping in mind that no meaningful \textit{local} order parameter can be defined. 

In this paper, we study an analogous order parameter, but with a different interpretation. We show that the non-local order can be interpreted in terms of the response to monopole configurations of an external $2+1D$ electromagnetic field~\cite{Henneaux1986,Pisarski1986,Diamantini1993,Fradkin1991, Grigorio2011,affleck1989}.
Importantly, basing our construction on such a response does not rely on any statistical magnetic field cancelling a physical one, and in fact it does not rely on any microscopic properties of the system. In particular, it does not depend on whether the system is composed of electrons or some other, more exotic state realizing a quantized Hall conductance, for instance a system with interacting bosons. This generality implies that our construction also works for Chern insulators, which do not require any external magnetic fields, and it could also be generalized to other topological states.

Besides the theoretical appeal of understanding the Chern insulator
states as a ``non-local order'' there is also a practical advantage:
an order parameter is often very easy to calculate with efficient
numerical algorithms, e.g. auxiliary field Monte Carlo. Hence, our
construction could be very useful in numerical studies of interacting
systems.

The remainder of this paper is organized as follows. In Sec.~\ref{sec:Orderparameter}, we
introduce the concept of non-local order parameters in the context
of Chern insulators. We then show in Sec.~\ref{sec:Monopole} how these order parameters can be interpreted as insertions of $2+1D$ $U(1)$ monopoles,
and in Sec.~\ref{sec:MonopoleResponse} how the order parameters can be obtained from an effective response theory of such insertions. In Secs.~\ref{sec:TrivialState} and~\ref{sec:CSterm}, we calculate the monopole response of the ground state for Chern insulator states with Hall conductance $\sigma_{H}=\frac{e^{2}}{h}$ and $\sigma_{H}=0$ respectively. In Sec.~\ref{generalization} we outline generalizations to other states, with $3+1D$ states characterised by the quantized magneto-electric effect as an example. We complement our analytical calculations with numerical modelling in Sec.~\ref{sec:Numerics} and summarize the paper in Sec.~\ref{sec:Conclusions} together with a few prospects towards future research.

\section{An order-parameter for Chern insulators}
\label{sec:Orderparameter}
As our starting point, we consider the non-local order parameter for the IQH effect. If we take the operator $\psi^{\dagger}(\mathbf{x})$,  which creates an electron at position $\mathbf{x}=(x,y)$,  and multiply it with an operator $\eta(\mathbf{x})$
\begin{equation}
\phi^{\dagger}(\mathbf{x})=\eta(\mathbf{x})\psi^{\dagger}(\mathbf{x})\ ,\label{eq:comnposite-boson}
\end{equation}
\begin{equation}
\eta(\mathbf{x})=e^{i\int d^{2}x^{\prime}\psi^{\dagger}(\mathbf{x}^{\prime})\psi(\mathbf{x}^{\prime})\arg(\mathbf{x}-\mathbf{x}^{\prime})},
\end{equation}
the resulting operator $\phi^{\dag}(\mathbf{x})$ is bosonic. We use the notation $\arg(\mathbf x)$ to denote the polar angle of $\mathbf x$.

It can be shown analytically~\cite{Girvin1987} that this operator has algebraically decaying correlation functions in $\Ket{LLL}$, the state of the filled lowest Landau level:
when $|\mathbf{x}-\mathbf{x}^{\prime}|$ is large compared to the
magnetic length ($l=\sqrt{eB/\hbar}$, where $B$ is the external magnetic field), the following expression holds:
\begin{equation}
\Braket{LLL|\phi(\mathbf{x})\phi^{\dagger}(\mathbf{\mathbf{x}^{\prime}})|LLL}\propto|\mathbf{x}-\mathbf{x}^{\prime}|^{-1/2}\ .\label{eq:ODLRO}
\end{equation}
For a local (product) state, however, the analogous correlator vanishes
as long as $\mathbf{x}\neq\mathbf{x}^{\prime}$. This suggests that the
algebraic off diagonal long range order (ODLRO) \eqref{eq:ODLRO}
can be used to distinguish between a trivial insulator and a IQH state. The idea of studying this operator
is three decades old~\cite{Girvin1987,Read1989,Sondhi1994,Sondhi1995} and was further suggested as a way to understand the Laughlin state~\cite{Laughlin1983}, i.e. a fractional quantum Hall state. We will here view it from a different perspective and show that the long range order is a characteristic property of the Hall conductance $\sigma_H$. For each integer $\n$ in 
\begin{equation}
\sigma_{H}=\n\frac{e^{2}}{h} \label{eq:hall-conductance},
\end{equation}
there exists an associated operator which characterize the corresponding state. We show that algebraic ODLRO is directly connected with singular behavior in the presence of a $U(1)$ monopole (or the absence of such behavior, in the case of the $\sigma_{H}=0$ state). For each value $\n$ above, the operator $\left.\phi_{m}(\mathbf{x})\right|_{m=\n}$  distinguishes between
the different values of $\n$ in the following way:
if $\Ket{\Psi_{\n}}$ is an insulating state with $\sigma_{H}=\n e^{2}/h$, the following asymptotic behaviors hold
\begin{subequations}
\label{eq:assymbehaviour}
\begin{align}
\Braket{\Psi_{\n}|\phi_{m}(\mathbf{x}^{\prime})\phi_{m}^{\dagger}(\mathbf{x})|\Psi_{\n}} & \propto|\mathbf{x}-\mathbf{x}^{\prime}|^{-\alpha}\quad m=\n\\
\Braket{\Psi_{\n}|\phi_{m}(\mathbf{x}^{\prime})\phi_{m}^{\dagger}(\mathbf{x})|\Psi_{\n}} & <e^{-|\mathbf{x}-\mathbf{x}^{\prime}|/\lambda}\hspace{1.2em}m\neq \n
\end{align}
\end{subequations}
with increasing $|\mathbf{x}-\mathbf{x}^{\prime}|$ (here, $\alpha$ and
$\lambda$ are non-universal constants). What are then the expressions
for $\phi_{m}^{}(\mathbf{x})$ or $\phi_{m}^\dagger(\mathbf{x})$? It turns out that the important component is $\eta(\mathbf{x})$ in Eq.~\eqref{eq:comnposite-boson}, which also defines $\phi_{0}^\dagger(\mathbf{x})$. The precise form of $\phi_{m}^{\dagger}(\mathbf{x})$ is in fact not important, but only the property
\begin{equation}
\phi_{m}^{\dagger}(\mathbf{x})=\chi\eta(\mathbf{x}),
\end{equation}
with $\chi$ being some product of creation operators satisfying
\begin{equation}
\int_{R_{\mathbf{x}}^{\epsilon}}d^{2}x^{\prime}[n(\mathbf{x}^{\prime}),\chi]=m\chi\ ,\label{eq:definingprop}
\end{equation}
where $n(\mathbf{x}^{\prime})$ is the number density of the $U(1)$
charge, and $R_{\mathbf{x}}^{\epsilon}$ is a region centered at $\mathbf{x}$
with a radius $\epsilon$ small compared to $\lambda$ in Eq.~\eqref{eq:assymbehaviour}. If the $U(1)$ charge density operator is given by 
\begin{equation}
n(\mathbf{x})=\psi^{\dagger}(\mathbf{x})\psi(\mathbf{x})\ ,\label{eq:u1density}
\end{equation}
we can e.g. take
\begin{align}
\label{eq:expliciteform}
&\phi_{1}^{\dagger}(\mathbf{x}) =\psi^{\dagger}(\mathbf{x})\eta(\mathbf{x}), \\
&\phi_{-1}^{\dagger}(\mathbf{x}) =\psi(\mathbf{x})\eta(\mathbf{x})\ .
\end{align}
The precise form of $\eta(\mathbf{x})$ is in fact not crucial either. The important feature is only that it can be interpreted as the insertion of a $2+1D$ $U(1)$ monopole. To illustrate this statement, and the asymptotic behavior in~Eq.~\eqref{eq:assymbehaviour}, we next take
a closer look at the operator $\eta(\mathbf{x})$.

\section{$\eta$ as a 2+1D monopole}
\label{sec:Monopole}
In this section, we show that $\eta(\mathbf{x})$ can be interpreted as the time-evolution operator associated with the insertion of a $2+1D$ $U(1)$ monopole. To avoid confusion, we begin by explaining what we mean with a monopole in various space-time dimensions. Our convention is that $\hbar=1$ and we absorb the electrical charge in the definition of the electromagnetic vector potential
$A_{\mu}$ and similarly in the electromagnetic field tensor $F_{\mu\nu}=\partial_{\mu}A_{\nu}-\partial_{\nu}A_{\mu}$. We denote space-time and space indices by greek and latin letters respectively and repeated indices are to be summed over. 

The electric source current density $j^{\mu}$, i.e. the $U(1)$ number current, is defined by the Maxwell equation 
\begin{equation}
\partial_{\mu}F^{\mu\nu}=j^{\mu}.
\end{equation}
Similarly, we may define a monopole source current, $j_{m.p.}$ as the right hand side in the dual Maxwell equation 
\begin{equation}
\partial_{\mu}\tilde{F}^{\mu\nu_{1}\dots\nu_{D-3}}=j_{m.p.}^{\nu_{1}\dots\nu_{D-3}}\label{dualeq}
\end{equation}
where the dual field strength is defined by 
\begin{equation}
\tilde{F}^{\sigma_{1}\dots\sigma_{D-2}}=\epsilon^{\sigma_{1}\dots\sigma_{D-2}\mu\nu}F_{\mu\nu},
\end{equation}
in which $\varepsilon^{\lambda\sigma\mu\dots\nu}$ is the totally anti-symmetric Levi-Civita tensor. It is crucial to  note that only in $3+1D$ is the elementary monopole source a current corresponding to a point particle. In $4+1D$, it is the world sheet of a string and in $2+1D$ it is a point in space-time often (in Euclidian space) referred to as an instanton. It is the latter case that is the main focus in this paper, but the higher dimensional case will be relevant for constructing order parameters for $3+1D$ $U(1)$ states~\cite{Fu2007,Moore2007,Roy2009,Ryu2010,Hosur2010,Chan2016}.  
We therefore continue by discussing only the $2+1D$ monopole in more detail. In this case, Eq.~\eqref{dualeq} reads
\begin{equation}
\varepsilon^{\sigma\mu\nu}\partial_{\sigma}F_{\mu\nu}=\partial_{t}B-\epsilon^{ij}\partial_{i}E_{j}=j_{m.p.}=\delta(t)\delta^{2}(\mathbf{x}-\mathbf{x}^{\prime}),\label{dual3}
\end{equation}
where we have taken the space-time monopole density to be a delta function of unit
(magnetic) charge. Clearly, there are many possible field configurations
that satisfy Eq.~\eqref{dual3}. The one we shall make use of reads (the reason will become clear in the next section)
\begin{equation}
\vec{E}(\mathbf{x},t)=-\frac{1}{2\pi}\delta(t)\vec{\nabla}\arg(\mathbf{x}-\mathbf{x}^{\prime})\ ,\label{realize}
\end{equation}
, which amounts to creating an instantaneous pulse of electric field.
Fixing a gauge, we can obtain Eq.~\eqref{realize} from the gauge potential
\begin{align}
A^{0}&=\frac{1}{2\pi}\delta(t)\arg(\mathbf{x}-\mathbf{x}^{\prime})& \vec A&=0
\,.\label{scalarpot}
\end{align}
Assuming further a Hamiltonian
\begin{align}
H=\tilde{H}(\vec{A})+\int d^{2}x\,n(\mathbf{x})A^{0}(\mathbf{x}),
\end{align}
where $\tilde{H}(\vec{A}(t))$ is the non-singular part and $n(\mathbf{x})$ is the local charge density, we can calculate the infinitesimal time-evolution operator $U(-\epsilon,\epsilon)$ from time $t=-\epsilon$ to time $t=\epsilon$ as the time ordered (denoted by $\mathcal{T}$) exponential of the Hamiltonian according to
\begin{align}
\label{eq:UEv}
&\lim_{\epsilon\rightarrow 0}U(-\epsilon,\epsilon)=\notag\\&\lim_{\epsilon\rightarrow0}\mathcal{T}\exp\Biggl({i\int d^{2}x\int_{-\epsilon}^{\epsilon}dt\delta(t)\arg(\mathbf{x}-\mathbf{x}^{\prime})n(\mathbf{x)}}\notag\\
&+i\int_{-\epsilon}^{\epsilon}dt\tilde{H}(\vec{A}(t))\Biggr)=e^{i\int d^{2}x\arg(\mathbf{x}-\mathbf{x}^{\prime})n(\mathbf{x)}}=\eta(\mathbf{x}),
\end{align}
where we used that the contribution from $\tilde{H}(\vec{A}(t))$ vanishes as $\epsilon\rightarrow 0$. We conclude from Eq.~\eqref{eq:UEv} that the application of the operator $\eta(\mathbf{x})$ can be interpreted as an infinitesimal time-evolution of the ground state in the presence of a monopole source insertion. We next show how this result allows us to obtain the order parameters $\phi_m(\mathbf{x})$ from $U(1)$ response theory. 

\section{U(1)-response theory and Monopole response}
\label{sec:MonopoleResponse}
Here, we show how the algebraic ODLRO follows from the $U(1)$
response theory of Chern-insulator states. To proceed in a self-contained manner, we begin by recalling some general formalism.

Let $\mathcal{W}[A_{\mu}]$ denote the functional of an external $U(1)$ gauge field, which generates connected current density correlation functions. In a state given by the density matrix $\rho$, the functional is given by
\begin{align}
&\mathcal{W}[A_{\mu}]=-i\log\mathcal{Z}[A_{\mu}]\equiv-i\log\text{Tr}\left(\mathcal{T}e^{i\int dtH(A_{\mu}(t))}\rho\right).\label{eq:functionaldefW}
\end{align}
To see that this is a valid expression, remember the definition
\begin{equation}
J^{\mu}(\mathbf{x})=\left.\frac{\delta H(A_{\mu})}{\delta A_{\mu}}\right|_{A=A^{bg.}}
\end{equation}
where $A^{bg}$ is a background field. Using
\begin{equation}
J^{\mu}(\mathbf{x},t)=\mathcal{T}e^{-i\int^{t}dt'H(A_{\mu}^{bg}(t'))}J^{\mu}(\mathbf{x},0)\mathcal{T}e^{i\int^{t}dt'H(A_{\mu}^{bg}(t'))},
\end{equation}
one directly gets
\begin{equation}
\left.\frac{\delta\mathcal{W}[A_{\mu}]}{\delta A_{\mu}}\right|_{A=A^{bg.}}=\Braket{J^{\mu}(\mathbf{x},t)}\equiv\text{Tr}\left(J^{\mu}(\mathbf{x},t)\rho\right).\label{currexp}
\end{equation}
The second order functional derivative gives the co-variance (or connected
two-point function) of the current-densities at different points (or
variance if the derivatives are evaluated at the same point) 
\begin{align}
\label{eq:Covariance}
&\left.\frac{\delta^{2}\mathcal{W}[A_{\mu}]}{\delta A_{\mu}(\mathbf{x}_{1},t_{1})\delta A_{\nu}(\mathbf{x}_{2},t_{2})}\right|_{A=A^{bg.}}=\notag\\
&\text{Cov}(J^{\nu}(\mathbf{x}_{2},t_{2})J^{\nu}(\mathbf{x}_{2},t_{2}))
\equiv\text{Tr}\bigg(J^{\nu}(\mathbf{x}_{2},t_{2})J^{\nu}(\mathbf{x}_{2},t_{2})\rho\bigg)\notag\\
&-\text{Tr}\bigg(J^{\mu}(\mathbf{x}_{1},t_{1})\rho\bigg)\text{Tr}\bigg(J^{\nu}(\mathbf{x}_{2},t_{2})\rho\bigg),
\end{align}
and analogously for multiple current insertions.

We assume that $\rho$ is the ground state of the gapped Hamiltonian $H(A_{\mu}^{bg}(t=-\infty))$ and that $H(A_{\mu}^{bg}(t))$ is gapped for all time. Then, it follows from the adiabatic theorem that $\mathcal{W}[A_{\mu}]$
is local in time for fields $A_{\mu}$ varying sufficiently slowly.
If the Hamiltonian is local, $\mathcal{W}[A_{\mu}]$ will also be
local in space, and because of current conservation it is also gauge
invariant.  Assuming (approximate) translation and rotation symmetry over long distances we can then expand the functional as follows 
\begin{align}
&\mathcal{W}[A_{\mu}]=-\frac{1}{4\pi}\int dtd^{2}x\Biggl(\n\varepsilon^{\mu\nu\lambda}A_{\mu}\partial_{\nu}A_{\lambda}\notag\\
&+\alpha\vec{E}_{A}^{2}-\beta B_{A}^2\biggr)+\cdots\label{eq:response}
\end{align}
where $\vec{E}_{A}$ and $B_{A}$ are the electric and magnetic fields
associated with $A$ and the dots denote higher order terms which
dictate the short distance and strong field dependence. Since we are interested in the opposite limit, these terms will from now on be neglected.

The integer $\n$ in Eq.~\eqref{eq:response} has a topological significance. In an IQH state it is simply the number of filled Landau levels, and more generally in a Chern insulator it is the first Chern number~\cite{Thouless1982,Kohomoto1985,Niu1985,Haldane1988}. Using Eq.~\eqref{currexp}, the Chern-Simons term in Eq.~\eqref{eq:response} yields, as units are restored, Eq.~\eqref{eq:hall-conductance} which demonstrates the topological nature of the Hall conductance. Since the integer $\n$ cannot change continuously, each integer defines a distinct state of a gapped zero temperature system, i.e. Chern-insulator states or IQH states, depending on the context.

Our claim is now that the correlation functions in Eq.~\eqref{eq:assymbehaviour} can be inferred directly from the response action \eqref{eq:response}, although Eq.~\eqref{eq:response} is formally only valid for slowly varying, smooth configurations of the background $U(1)$ gauge field. In the following we will only make use of the defining property, Eq.~\eqref{eq:definingprop}, of $\phi_{m}^{\dagger}(\mathbf{x})$ and the analysis is valid for all non-zero values of $m$. For simplicity, we consider only
$\phi_{1}^{\dagger}(\mathbf{x})$ and assume that the charge density is composed only of a single species of fermions, $n(\mathbf{x})=\psi^{\dagger}(\mathbf{x})\psi(\mathbf{x})$.
We also take the explicit form $\phi_{1}^{\dagger}(\mathbf{x})=\psi^{\dagger}(\mathbf{x})\eta(\mathbf{x})$ and use it to show the main claim of the paper, that the asymptotic behavior \eqref{eq:assymbehaviour} follows from the long-range $U(1)$ response. 

Now to the question of why we choose the particular electric and magnetic field configurations of Eq.~\eqref{realize} as a solution to the monopole equation in Eq.~\eqref{dual3}. Our aim is to arrive at the expression for a creation operator of an order parameter field. The field configuration should therefore be instantaneous in time. However, we could have chosen any electric field given by a gauge potential of the form 
\begin{equation}
A^{0}=\frac{1}{2\pi}\delta(t)\arg(\mathbf{x}-\mathbf{x}^{\prime})+f(\mathbf x)\, ,
\end{equation}
where $f$ is any singled valued function. As we will describe in more detail below, a Chern-Simons term implies that a monopole will be associated with electrical charge, which is the reason for the difference in the behavior of the non-local order parameter in the different states. 

We, therefore, do not want any charge to be created by the Maxwell term 
\begin{equation}
\int dtd^{2}x\,\left(\alpha\vec{E}_{A}^{2}-\beta B_{A}^2\right).
\end{equation}
Consequently, $A^0$ should correspond to the solution to the equation of motion obtained by varying the Maxwell term with respect to $A^0$. That is 
\begin{equation}
	\nabla^2 A^0=\nabla^2 f=0 \label{fcondition}.
\end{equation}
With the boundary condition $f\rightarrow0$ when $\mathbf x\rightarrow\infty$ we get $f=0$.

This analysis also hints why it is reasonable to derive the ODLRO only from the low order correlation function. If we consider also higher-order translationally and rotationally invariant terms in the response action, i.e. terms of the form
\begin{equation}
\int dtd^{2}x\, (\vec E_A^2)^m (B_A^{2})^n\, 
\end{equation}
where $m$ and $n$ are integers, the field configuration \eqref{scalarpot} is still a solution to the equation one obtains by varying these terms with respect to $A^0$. With this background, we present in the following two sections the derivation of Eq.~\eqref{eq:assymbehaviour}. 

\section{The trivial Chern insulator state}
\label{sec:TrivialState}
We first consider the topologically trivial $\n=0$ state, where we
want to prove 
\begin{align}
&\Braket{\Psi_{0}|\phi_{1}(\mathbf{x}^{\prime})\phi_{1}^{\dagger}(\mathbf{x})|\Psi_{0}}=\notag\\
&=-\Braket{\Psi_{0}|\psi(\mathbf{x}^{\prime})\psi^{\dagger}(\mathbf{x})\eta^{\dagger}(\mathbf{x}^{\prime})\eta(\mathbf{x})|\Psi_{0}}=0\ \text{if}\ \mathbf{x}\neq\mathbf{x}^{\prime}.\label{eq:corr1}
\end{align}
To this end, we compute the variance of the total charge $N_{\mathbf{x}}^{\epsilon}$ in a small region $R_{\mathbf{x}}^{\epsilon}$, in the state 
\begin{equation}
\Ket{\eta^{\dagger}\eta\Psi_{0}}\equiv\eta^{\dagger}(\mathbf{x}^{\prime})\eta(\mathbf{x})\Ket{\Psi_{0}}\ .
\end{equation}
It follows from Sec.~\ref{sec:Monopole} that acting with the operators $\eta^{\dagger}(\mathbf{x}^{\prime})\eta(\mathbf{x})$
is equivalent to a perturbation with the background field $A_{b.g.}=A^{\mathbf{x}^{\prime}\,\mathbf{x}}$
where 
\begin{equation}
A_{0}^{\mathbf{x}^{\prime}\,\mathbf{x}}(\mathbf{x}^{\prime\prime})=\frac{1}{2\pi}\delta(t)\arg(\mathbf{x}^{\prime\prime}-\mathbf{x})-\frac{1}{2\pi}\delta(t)\arg(\mathbf{x}^{\prime\prime}-\mathbf{x}^{\prime})\ .
\end{equation}
Using the response functional \eqref{eq:response} with $\n=0$ in Eq.~\eqref{eq:Covariance}, we obtain
\begin{equation}
\text{var}(N_{\mathbf{x}}^{\epsilon})=-\frac{\alpha}{2\pi}\int_{R_{\mathbf{x}}^{\epsilon}\times R_{\mathbf{x}}^{\epsilon}}\hspace{-25bp}d^{2}x^{\prime}d^{2}x^{\prime\prime}\nabla^{2}\delta^{2}(\mathbf{x}^{\prime}-\mathbf{x}^{\prime\prime})=0.
\end{equation}
In a similar way, we also calculate the expectation value of $N_{\mathbf{x}}^{\epsilon}$  and we conclude that if $\mathbf{x}^{\prime}\notin R_{\mathbf{x}}^{\epsilon}$
it follows that $\psi^{\dagger}(\mathbf{x})\Ket{\eta^{\dagger}\eta\Psi_{0}}$
and $\psi^{\dagger}(\mathbf{x}^{\prime})\Ket{\Psi_{0}}$ are eigenstates
of $N_{\mathbf{x}}^{\epsilon}$ with different eigenvalues. Since
$\epsilon$ can be made arbitrarily small, Eq.~\eqref{eq:corr1} follows.
One may now wonder why the right-hand side of Eq.~\eqref{eq:corr1} is zero and not an exponential as claimed in Eq.~\eqref{eq:assymbehaviour}. This is an artifact of using the fully local response functional (which is the case when $\n=0$):
\begin{align}
\mathcal{W}[A_{\mu}]&=\int dtd^{2}x\,\left(\alpha\vec{E}_{A}^{2}-\beta B_{A}^2\right)\notag\\
&=\int dtd^{2}xd^{2}x^{\prime}\delta^{2}(\mathbf{x}-\mathbf{x}^{\prime})\Bigl(\alpha\vec{E}_{A}(\mathbf{x}^{\prime})\cdot\vec{E}_{A}(\mathbf{x})\notag\\
&-\beta B_{A}(\mathbf{x}^{\prime})\cdot B_{A}(\mathbf{x})\Bigr). 
\end{align}
This expression is only valid for calculating long distance
properties of the system. 
To illustrate this point, let us see what happens with our previous argument when we regularize the delta function in the equation above as $\delta^{2}(\mathbf{x}-\mathbf{x}^{\prime})=(\sqrt{\pi}\delta)^{-2}e^{-|\mathbf{x}-\mathbf{x}^{\prime}|^{2}/\delta^{2}}$.
The expectation value $N_{\mathbf{x}}^{\epsilon}$ remains the same, but the variance becomes instead
\begin{equation}
\text{var}(N_{\mathbf{x}}^{\epsilon})=\frac{2\alpha\varepsilon^{4}}{\delta^{4}}e^{-\varepsilon^{2}/\delta^{2}}\ .
\label{eq:shortdistvar}
\end{equation}
By using that the $N_{\mathbf{x}}^{\epsilon}$ expectation values
in the states $\psi^{\dagger}(\mathbf{x})\Ket{\eta^{\dagger}\eta\Psi_{0}}$ and $\psi^{\dagger}(\mathbf{x}^{\prime})\Ket{\Psi_{0}}$ differ
by $1$ (since $\psi^\dagger(\mathbf x)$ creates a unit charge at $\mathbf x$), and the expression \eqref{eq:shortdistvar} for the variance, we
can put an upper bound on the correlation function 
\begin{equation}
|\braket{\Psi_{0}|\phi_{1}(\mathbf{x}^{\prime})\phi_{1}^{\dagger}(\mathbf{x})|\Psi_{0}}|^{2}\leq\alpha\frac{|\mathbf{x}-\mathbf{x}^{\prime}|^{4}}{\delta^{4}}e^{-|\mathbf{x}-\mathbf{x}^{\prime}|^{2}/\delta^{2}}\ .
\end{equation}
This argument shows that the scale $\lambda$ in Eq.~\eqref{eq:assymbehaviour}
is lost in the local description $\delta\rightarrow0$, i.e.,
with the microscopic scales set to zero.

We next turn to the correlator
\begin{equation}
\Braket{\Psi_{0}|\phi_{0}(\mathbf{x}^{\prime})\phi_{0}^{\dagger}(\mathbf{x})|\Psi_{0}}=\Braket{\Psi_{0}|\eta(\mathbf{x}^{\prime})\eta^{\dagger}(\mathbf{x})|\Psi_{0}}\ .
\end{equation}
From the definition \eqref{eq:functionaldefW} for the response functional
we can identify this expression with $e^{-i\mathcal{W}[A^{\mathbf{x}^{\prime}\,\mathbf{x}}]}$. We will make extensive use of this indentifcation later, but for now it suffices to use the correlation
function directly from the definition 
\begin{equation}
\label{eq:etaDef2}
\eta(\mathbf{x})=e^{i\int d^{2}x^{\prime}n(\mathbf{x}^{\prime})\arg(\mathbf{x}-\mathbf{x}^{\prime})}\ .
\end{equation}
This follows because the delta functions $\delta(t)$ in $A^{\mathbf{x}^{\prime}\,\mathbf{x}}$
must be regularized to evaluate $e^{-i\mathcal{W}[A^{\mathbf{x}^{\prime}\,\mathbf{x}}]}$, but by using Eq.~\eqref{eq:etaDef2}, we avoid that complication. 

To compute the expectation value of the integral of an
operator we can,  since $\mathcal{W}[A_{\mu}]$ is quadratic, use the
cumulant expansion
\begin{equation}
\langle e^{\Omega}\rangle=\exp\left(\langle\Omega\rangle+\frac{1}{2}(\langle\Omega^{2}\rangle-\langle\Omega\rangle^{2})\right).
\end{equation}
We then obtain
\begin{align}
\label{eq:cumulant2}
 & \log\Braket{\Psi_{0}|\eta(\mathbf{x}^{\prime})\eta^{\dagger}(\mathbf{x})|\Psi_{0}}\nonumber \\
 & =-\frac{\alpha}{4\pi}\int d^{2}x^{\prime\prime}\left(\vec{\nabla}\arg(\mathbf{x^{\prime\prime}}-\mathbf{x}^ {})-\vec{\nabla}\arg(\mathbf{x^{\prime\prime}}-\mathbf{x}^{\prime})\right)^{2}.
\end{align}
To evaluate this particular spatial integral, we use without loss
of generality, polar coordinates $\{r,\varphi\}$ with the origin
at $\mathbf{x^{\prime}}$. We find 
\begin{align}
\label{eq:polarintegral}
 & \int d^{2}x^{\prime\prime}\left(\vec{\nabla}\arg(\mathbf{x^{\prime\prime}}-\mathbf{x}^ {})-\vec{\nabla}\arg(\mathbf{x^{\prime\prime}}-\mathbf{x}^{\prime})\right)^{2}\nonumber \\
 & =\int_{0}^{\infty}drr\frac{2\pi\text{R}^{2}}{r^{2}(r+R)\left|r-R\right|}\nonumber \\
 & =-\int_{\delta}^{R-\delta}dr\frac{2\pi\text{R}^{2}}{r(r+R)(r-R)}+\nonumber \\
 & +\int_{R+\delta}^{\infty}dr\frac{2\pi\text{R}^{2}}{r(r+R)(r-R)}=4\pi\log(R)+\ldots,
\end{align}
where $R\equiv|\mathbf{x}-\mathbf{x}^{\prime}|$ and the ellipsis
denote $R$-independent terms which diverge as the short distance
cutoff $\delta\rightarrow0$. From Eqs.~\eqref{eq:cumulant2} and~\eqref{eq:polarintegral}, we conclude that 
\begin{equation}
\Braket{\Psi_{0}|\phi_{0}(\mathbf{x}^{\prime})\phi_{0}^{\dagger}(\mathbf{x})|\Psi_{0}}\propto|\mathbf{x}-\mathbf{x}^{\prime}|^{-\alpha}
\end{equation}
where $\alpha$ is the very same constant as in the response functional~\eqref{eq:response}. Thereby, we obtained the previously
stated algebraic decay, but we also see that the decay is intimately related to the system's electric polarization. We next turn to the non-trivial Chern-insulator states. 
\section{Chern insulators---importance of the Chern-Simons term}
\label{sec:CSterm}
In this section, we consider the states $\Ket{\Psi_{\n}}$ with $\n\neq0$. To make the argumentation more concrete, we focus on $\Ket{\Psi_{1}}$.
We begin by proving 
\begin{equation}
\Braket{\Psi_{1}|\eta(\mathbf{x}^{\prime})\eta^{\dagger}(\mathbf{x})|\Psi_{1}}=0\ .
\label{eq:CSmakezero}
\end{equation}
As in the previous section we can use the identity 
\begin{equation}
\Braket{\Psi_{1}|\eta(\mathbf{x}^{\prime})\eta^{\dagger}(\mathbf{x})|\Psi_{1}}=e^{-i\mathcal{W}[A_{\mu}^{\mathbf{x}^{\prime}\,\mathbf{x}}]}\ .\label{eq:corrresp}
\end{equation}
Let us factor the response functional according to
\begin{equation}
e^{-i\mathcal{W}[A_{\mu}]}=e^{-i\mathcal{W}_{CS}[A_{\mu}]}e^{-i\mathcal{W}_{Max.}[A_{\mu}]}\label{eq:factorization},
\end{equation}
with 
\begin{align}
&\mathcal{W}_{CS}[A_{\mu}] =-\frac{1}{4\pi}\int dtd^{2}x\,\varepsilon^{\mu\nu\lambda}A_{\mu}\partial_{\nu}A_{\lambda}\label{eq:responseCS},\\
&\mathcal{W}_{Max.}[A_{\mu}] =-\frac{1}{4\pi}\int dtd^{2}x\,\left(\alpha\vec{E}_{A}^{2}-\beta B_{A}^2\right)\ .
\end{align}
It turns out that it is the Chern-Simons term that is responsible for the zero result in Eq.~\eqref{eq:CSmakezero}, so we temporarily neglect the the Maxwell term.

The electrical field $\vec{E}(\mathbf{x}^{\prime\prime})$, defining $A_{\mu}^{\mathbf{x}^{\prime}\,\mathbf{x}}$,
is singular at $t=0$, $\mathbf{\mathbf{x}^{\prime\prime}}=\mathbf{x}$
or $\mathbf{\mathbf{x}^{\prime\prime}}=\mathbf{x}^{\prime}$. This feature
was not a problem in the previous section, where we used a direct
regularization by removing the singularity by cutting away a small
region in space-time, around the singularities and took the limit
of the region becoming infinitesimally small. The problem with this procedure in the presence of a Chern-Simons term is that the presence of space-time boundaries breaks gauge invariance. We cannot therefore simply cut away a small region for the non-trivial states. We can however overcome this limitation by rewriting the expression \eqref{eq:responseCS} in a way that is equivalent when treated in a system without a boundary but remains valid even if a boundary is present. We use use the ``functional boson'' identity~\cite{Chan2013} and write
\begin{align}
\label{eq:funcboson}
& e^{-i\mathcal{W}_{CS}[A_{\mu}]}=\int D[b_{\mu}]e^{i S[A_\mu,b_\mu]}, \notag \\
& S[A_\mu,b_\mu] = \int dtd^{2}x\,\varepsilon^{\mu\nu\lambda}b_{\mu}\partial_{\nu}A_{\lambda}+\int dtd^{2}x\,\varepsilon^{\mu\nu\lambda}b_{\mu}\partial_{\nu}b_{\lambda},
\end{align}
which clearly holds when space-time does not have a boundary. To proceed, we simply define our Chern-Simons response functional as in Eq~\eqref{eq:funcboson} also in the presence of boundaries. When space-time does have a boundary, there is no gauge invariance for $b_{\mu}$ on the boundary and these ``would-be'' gauge degrees of freedom can be written in terms of a scalar boson $\Theta$ (see e.g., Ref. \onlinecite{Wen1995}). Let us next imagine removing a small ball $B$ from space-time $\mathcal{M}$, leaving us with the new space-time $\mathcal{M}/B$. When we next integrate out $b_{\mu}$, we find
\begin{align}
e^{-i\mathcal{W}_{CS}[A_{\mu}]}=&\exp\left(\frac{i}{4\pi}\int_{\mathcal{M}/B}\hspace{-1.5em}dtd^{2}x\,\varepsilon^{\mu\nu\lambda}A_{\mu}\partial_{\nu}A_{\lambda}\right)\notag\\
&\times\int D[\Theta]e^{i\int_{\partial B}d^{2}x\,\mathcal{L}[A_{\mu},\Theta]}.\label{eq:chiral-boson}
\end{align}
For our purpose, there is no need to fully specify the boson
Lagrangian $\mathcal{L}[A_{\mu},\Theta]$.  The only property we need is
that it contains an ``anomalous'' term 
\begin{equation}
\mathcal{L}_{\text{anom.}}=\Theta N_{\sigma}\varepsilon^{\sigma\mu\nu}\partial_{\mu}A_{\nu}\ ,
\end{equation}
where $N_{\sigma}$ is the normal to the boundary $\partial B$ of
the ball $B$. If $\partial B$ surrounds a monopole, then
\begin{equation}
\int_{\partial B}d^{2}x\,N_{\sigma}\varepsilon^{\sigma\mu\nu}\partial_{\mu}A_{\nu}=s_{m},
\end{equation}
where $s_{m}$ is the monopole strength. Since the functional integral \eqref{eq:chiral-boson} contains an integral over the constant mode $\Theta_{const.}$, it follows that if there is a boundary surrounding a monopole with strength $s_{m}$, we must have
\begin{equation}
e^{i\mathcal{W}[A_{\mu}]}\propto\int D[\Theta_{const}]e^{i\Theta_{const}s_{m}}\propto\delta_{0,s_{m}}\ .
\end{equation}
We can therefore conclude that due to the Chern-Simons term
\begin{equation}
\Braket{\Psi_{1}|\eta(\mathbf{x}^{\prime})\eta^{\dagger}(\mathbf{x})|\Psi_{1}}=0.
\end{equation}
From these calculations, we can in fact draw additional conclusions.  When we compute current expectation values (away from the points $\mathbf x$ and $\mathbf x^\prime$) in states where $\eta(\mathbf x)$ or $\eta(\mathbf x^\prime)$ are inserted, the boson $\Theta$ does not contribute (other than possibly with the above calculated delta function). Because of current conservation, we can also compute the change of charge at e.g., $\mathbf x$ from time $t=-\epsilon$ to time $t=\epsilon$: this change is simply the current that has flown to $\mathbf x$ during that time. By a direct calculation, it follows that the electric field in Eq.~\eqref{realize} (corresponding to $\eta(\mathbf{x})$) changes the charge by $1$ if the Hall conductance takes the value corresponding to $\n=1$. Furthermore, an operator which increases the expectation value of the charge by $1$ must be proportional to $\psi^\dagger(\mathbf x)$ since we assumed $n(\mathbf x)=\psi^\dagger(\mathbf x)\psi(\mathbf x)$. The effect of $\eta(\mathbf x)$ in a state with a $\n=1$ CS term is therefore, up to a constant, the same as the effect of $\psi^\dagger(\mathbf x)\eta(\mathbf x)$ in a state with the same response, but with $\n=0$. 

It is instructive to see some of the steps in this argument spelled out in more detail. We do this by first proving that with $\n=1$, $\eta(\mathbf x)$ changes the expectation value of $N_{\mathbf{x}}^{\epsilon}$ by $1$, i.e.
\begin{equation}
\Delta\equiv\Braket{\Psi_{1}|\eta(\mathbf{x})N_{\mathbf{x}}^{\epsilon}\eta^{\dagger}(\mathbf{x})|\Psi_{1}}-\Braket{\Psi_{1}|N_{\mathbf{x}}^{\epsilon}|\Psi_{1}}=1\ .\label{eq:charge-change}
\end{equation}
If this identity holds, and if both $\Ket{\Psi_{1}}$
and $\eta^{\dagger}(\mathbf{x})\Ket{\Psi_{1}}$ have zero variance
in $N_{\mathbf{x}}^{\epsilon}$, it implies that 
\begin{equation}
\Braket{\Psi_{1}|\psi(\mathbf{x})\eta^{\dagger}(\mathbf{x})|\Psi_{1}}\neq0.
\end{equation}
Because of current conservation we can calculate this change in charge as the charge that has flowed in to $R^\epsilon_{\mathbf x}$ by the insertion of $\eta(\mathbf{x})$
\begin{align*}
\Delta& =\lim_{\delta\rightarrow0}\int_{-\delta}^{\delta}dt\int_{\partial R_{\mathbf{x}}^{\epsilon}}d^{2}x^{\prime}\,N_{i}\Braket{\Psi_{1}|\eta(\mathbf{x})j^{i}(\mathbf{x^{\prime}})\eta^{\dagger}(\mathbf{x})|\Psi_{1}}
\end{align*}
where $\partial R_{\mathbf{x}}^{\epsilon}$ is the boundary of the
region defining $N_{\mathbf{x}}^{\epsilon}$ and $N_{\mu}$ is its
normal. By definition, this expression equals 
\begin{equation}
\Delta=\lim_{\delta\rightarrow0}\int_{-\delta}^{\delta}dt\int_{\partial R_{\mathbf{x}}^{\epsilon}}d^{2}x^{\prime}\,N_{i}\left.\frac{\delta}{\delta A_{i}(\mathbf{x}^{\prime},t)}\mathcal{W}[A]\right|_{A=A^{\mathbf{x}}}\ ,
\end{equation}
with 
\begin{equation}
A_{0}^{\mathbf{x}}(\mathbf{x}^{\prime})=\frac{1}{2\pi}\delta(t)\arg(\mathbf{x}^{\prime}-\mathbf{x}).
\end{equation}
We first note that $\mathcal{W}_{Max.}[A_{\mu}]$ gives no contribution to $\Delta$, so we focus next on the contribution from $\mathcal{W}_{CS}[A_{\mu}]$, for which we can use Eq.~\eqref{eq:chiral-boson}. We can choose the ball $B$ arbitrarily small and in particular choose it such that it does not intersect $\partial R_{\mathbf{x}}^{\epsilon}$. Then the functional integral part of \eqref{eq:chiral-boson}
\begin{equation}
\int D[\Theta]e^{i\int_{\partial B}d^{2}x\,\mathcal{L}[A_{\mu},\Theta]}
\end{equation}
gives no contribution, and we get
\begin{equation}
\Delta=\lim_{\delta\rightarrow0}\int_{-\delta}^{\delta}dt\int_{\partial R_{\mathbf{x}}^{\epsilon}}d^{2}x^{\prime}\,\left.N_{i}\varepsilon^{ij}\partial_{j}A_{0}\right|_{A_{\mu}=A_{\mu}^{\mathbf{x}}}=1\ .
\end{equation}
We can therefore conclude that the Chern-Simons term results in $\eta(\mathbf{x})\sim\psi^{\dagger}(\mathbf{x})$.

The Maxwell term gives the same contribution as in the previous
section, so we can finally conclude 
\begin{equation}
\Braket{\Psi_{1}|\phi_{1}^{}(\mathbf{x}^{\prime})\phi_{1}^{\dagger}(\mathbf{x})|\Psi_{1}}\propto|\mathbf{x}-\mathbf{x}^{\prime}|^{-\alpha}\ .
\end{equation}
For the other order-parameters $\phi_{m}(\mathbf{x})$ with $m\neq1$ we get zero which can again
be shown by the fact that the below bra and ket states have different
$N_{\mathbf{x}}^{\epsilon}$ eigenvalues, 
\begin{align}
\Braket{\Psi_{1}|\phi_{m}^{}(\mathbf{x}^{\prime})\phi_{m}^{\dagger}(\mathbf{x})|\Psi_{1}} & =0 & \text{if } & m\neq1\ .
\end{align}
As for the trivial state, the reason we get zero and not exponential decay happens because the length scale which dictates the exponential behavior is set by a scale already assumed to be zero. We therefore end this section with an argument for the exponential decay based on confinement of monopole-antimonopole pairs. 

A remarkable and well-known property of the Chern-Simons term is that it attaches flux to charges, which makes it a useful model for anyonic statistics~\cite{Wilczek1982}. Equally remarkable, but not as well-known, is that the opposite mechanism is also at work: the insertion of a monopole-antimonopole pair into the Maxwell-Chern-Simons theory induces an electrical current which flows through the connecting Dirac string~\cite{Henneaux1986,Pisarski1986,Diamantini1993,Grigorio2009,Grigorio2011}. Because of its physical charge, the string becomes a physical entity and the pair interact through a linearly confining potential. In particular, isolated monopoles (which have infinite string lengths) are completely suppressed in the partition function. We can therefore in Eq.~\eqref{eq:assymbehaviour}  identify the constant $\lambda$ as the inverse string tension and the asymptotic exponential decay follows from the linear potential between the monopoles. This should be contrasted to the trivial $\n=0$ phase, where we showed above that the monopoles interact through the ordinary $2+1D$ logarithmic Coulomb potential.
This argument ends our analytical treatment of the Chern insulator states, and we proceed to generalize our construction to other topological states of matter. 

\section{Generalization to 3+1D}
\label{generalization}
In this section, we discuss how a similar notion of non-local order-parameter can be constructed for topological insulators characterized by electromagnetic response in form of $\theta$-terms~\cite{Fu2007,Moore2007,Roy2009,Ryu2010,Hosur2010,Chan2016}. As opposed to the Chern-Simons term, the $\theta$-term is only quantized if there is either time-reversal or chiral symmetry which is therefore always assumed in this section. 

We consider $3+1D$, where the $\theta$-term reads 
\begin{align}
\mathcal W_\theta[A_{\mu}]=\frac{\nu}{4\pi}\int d^{4}x\,\vec{E}_A\cdot\vec{B}_A\ .\label{thetaterm}
\end{align}
Just as the Chern-Simons term binds an electric charge to a flux-insertion, this term binds $\nu/2$ charges to a magnetic monopole~\cite{witten79}.

Let us follow the recipe from the last section and construct a non-local order-parameter $\phi_\nu(\mathbf x^\prime)$ for the state with a given $\nu$ in Eq.~\eqref{thetaterm}. Let us take $\nu=1$ as an example. There will then be half of an electric charge associated with every monopole, so an the order-parameter, $\phi^\dagger_{\nu=1}(\mathbf x^\prime)$, should be of the form 
\begin{align*}
	\phi^\dagger_1(\mathbf x^\prime)=\eta^2(\mathbf x^\prime)\psi^\dagger(\mathbf x^\prime)\ ,
\end{align*}
where $\eta^\dagger(\mathbf x^\prime)$ now correspond to a singular electromagnetic field associated with creating a monpole at $\mathbf x^\prime$. Let us consider such a field. We want the field strength to obey
\begin{equation}
\epsilon^{\mu\nu\sigma\lambda}\partial_\nu F_{\sigma\lambda}(\mathbf x)=
\frac{1}{2\pi}\begin{cases}
\delta^{t,\mu}\delta^3(\mathbf x-\mathbf x^\prime) &t>0,\\
0 & t<0,
\end{cases}
\end{equation}
which corresponds to a stationary monopole at $\mathbf x^\prime$  after $t=0$, and no monopole before $t=0$. From Stokes theorem, we see that this is not possible. The world-line of the monopole must be closed (or extend to infinity), which means that we may not write down a field corresponding to creating a lone monopole. In other words $\eta$ is not well defined on its own. Instead, we consider the combination $\eta^\dagger(\mathbf x^{\prime\prime})\eta(\mathbf x^\prime)$ which corresponds to an electromagnetic field strength instead obeying 
\begin{multline}
\epsilon^{\mu\nu\sigma\lambda}\partial_\nu F_{\sigma\lambda}(\mathbf x)=\\
\frac{1}{2\pi}\begin{cases}
\delta^{t,\mu}\delta^3(\mathbf x-\mathbf{x}^\prime)-\delta^{t,\mu}\delta^3(\mathbf x-\mathbf x^{\prime\prime}) &t>0,\\
0 & t<0\ .
\end{cases}
\end{multline}
Now we cannot get any further from the topological properties alone. Again we must, as previously, be guided by the next term in the derivative expansion. If we assume translation and rotation invariance on long length scales, this term is on Maxwell form
\begin{align}
\mathcal{W}_{Max.}[A_{\mu}]=-\frac{1}{4\pi}\int dtd^{3}x\,\left(\alpha\vec{E}_{A}^{2}-\beta \vec B_{A}^2\right)\ .
\end{align}
Using the minimization of the Maxwell term as a guiding principle, we can construct a suitable field configuration. This can be done as follows. First we choose the field strength after $t=0$, i.e., when there are two stationary monopoles. By rotation symmetry of the Maxwell term the minimzer is just the rotation symmetric monopole,
\begin{align}
	\vec B_A(\mathbf x)=\frac{1}{4\pi} \frac{\vec r^\prime}{|\mathbf x-\mathbf x^\prime|^2}-\frac{1}{4\pi} \frac{\vec r^\prime}{|\mathbf x-\mathbf x^{\prime\prime}|^2}\ ,\label{monopoles}
\end{align}
where $\vec r^\prime$ and $\vec r^{\prime\prime}$ are the radial unit vectors centered at $\vec r^\prime$ and $\vec r^{\prime\prime}$ respectively. 
Since world-lines are needed to be closed we also need a singular instantaneus monopole current at $t=0$, moving a monopole charge from $\mathbf x^\prime$ to $\mathbf x$. Again by symmetry, we can conclude that a minimum of the Maxwell term is attained if it follows a straight line between $\mathbf x^\prime$ and $\mathbf x^{\prime\prime}$. 
We choose our coordinates such that $\mathbf x^\prime=(0,0,R/2)$ and $\mathbf x^{\prime\prime}=(0,0,-R/2)$.  We thus have the equation 
\begin{multline}
\epsilon^{\mu\nu\sigma\lambda}\partial_\nu F_{\sigma\lambda}(\mathbf x)=\\
\frac{1}{2\pi}\begin{cases}
\delta^{t,\mu}\delta^3(\mathbf x-\mathbf x^\prime) &t>0,\\
	H(z-R/2)H(R/2-z)\delta^{z,\mu}\delta(t)\delta^2(\mathbf x_\perp)&t=0\\
0 & t<0,
\end{cases}
\end{multline}
where, $\mathbf{x}_{\perp}$ denote the component of $\mathbf x$ in the $xy$-plane and $H$ is the Heaviside step funtion,
\begin{equation}
H(x)=
\begin{cases}
0 & x<0\\
1 & x\leq 0 \ .
\end{cases}
\end{equation}
With the additional requirement that the field takes the form \eqref{monopoles} for $t>0$ we have the general solution for singular part of the field configuration at $t=0$ of the form $\vec E(\mathbf x,t)=\delta(t)\left(\vec E_0(\mathbf x,\mathbf x^\prime,\mathbf x^{\prime\prime})+\vec \nabla f(\mathbf x)\right)$. If we take $\mathbf x^\prime=(0,0,R/2)$ and $\mathbf x^{\prime\prime}=(0,0,-R/2)$ then $\vec E_0$ is given by
\begin{align}
&\vec E_0(\mathbf x,\mathbf x^\prime,\mathbf x^{\prime\prime})\notag \\&=\frac{1}{4\pi}\Biggl(\frac{z-R/2}{\sqrt{x^{2}+y^{2}+(z-R/2)^{2}}}\vec{\nabla}\arg(\mathbf{x}_{\perp})\notag\\
&-\frac{z+R/2}{\sqrt{x^{2}+y^{2}+(z+R/2)^{2}}}\vec{\nabla}\arg(\mathbf{x}_{\perp})\Biggr)\ .
\end{align}
Here, $\arg$ is the 2+1D argument function in the $xy$-plane. As in the 2D case we have an arbitrary gradient term $\vec \nabla f(\mathbf x)$. Just as then we conclude $f\equiv0$ from minimizing the Maxwell term. 
Again just as the 2+1D case we then define $\eta^\dagger(\mathbf x^{\prime\prime})\eta(\mathbf x^\prime)$ as the time-evolution operator of this field, during an infinitesimally small time interval around $t=0$. This procedure leaves us again with just the minimal coupling term at $t=0$ since the other terms in the Hamiltonian are finite and vanish in the limit of vanishing time-interval around $t=0$. 

However, this time we cannot choose a gauge where the vector potential only depends on the charge density, but we must instead use the full current operator. Since the current is conserved we can, using the Helmholtz decomposition, write the current operator as the curl of a vector field according to
\begin{align}
	\hat{\vec j}(\mathbf y)=\frac{1}{4\pi}\vec\nabla \times \int d^3y\frac{\vec\nabla_y\times \hat{\vec j}(\mathbf y)}{|\mathbf x-\mathbf y|}
\end{align}
If we integrate by parts the minimal coupling term $j^\mu A_\mu$, containing such a decomposition, we end up with
\begin{align}
&\eta^\dagger(\mathbf x^{\prime\prime})\eta(\mathbf x^\prime)=\lim_{\epsilon\rightarrow0}U(-\epsilon,\epsilon)\notag\\
&=\lim_{\epsilon\rightarrow0}\mathcal{T}\lbrace e^{i\int_{-\epsilon}^{\epsilon}dt\tilde{H}(\vec{A}(t))} e^{i\int_{-\epsilon}^{\epsilon}dt\int d^3 x\, j^\mu A_\mu}\rbrace\notag\\
&=\exp{\left(\frac{i}{4\pi}\int d^3 x\,\vec E_0(\mathbf x,\mathbf x^\prime,\mathbf x^{\prime\prime})\cdot\int d^3y\,\frac{\vec\nabla_y\times \hat{\vec j}(\mathbf y)}{|\mathbf x-\mathbf y|}\right)}.
\end{align}
This is a double integral over space, and it includes the current operator in contrast to only the density operator as for the Chern insulator. The operator is therefore much more involved to handle numerically, and we leave numerical evaluations for $3+1D$ to the future. However, if we are only interested in the order-parameter for the $\nu=0$ state there is a convenient way to simplify this expression. If we take $R\rightarrow \infty$, we get 
\begin{align}
	\vec E_0=\frac{1}{2\pi}\arg(\mathbf{x}_{\perp})\ .
\end{align}
The situation is now very similar to the Chern insulator and we can choose $\vec A=0$ to get 
\begin{align}
\label{eq:3Dsimple}
	& \lim_{R\rightarrow\infty}\phi_0^\dagger(\mathbf x^{\prime\prime})\phi_0(\mathbf x^{\prime\prime})	\equiv \lim_{R\rightarrow\infty}\eta^\dagger(\mathbf x^{\prime\prime})\eta(\mathbf x^\prime)\notag\\&=e^{i\int d^{3}x\arg(\mathbf{x}_\perp)n(\mathbf{x)}}.
\end{align}
This operator corresponds to sending one monopole to $z=+\infty$ and the other to $z=-\infty$. If $\nu$ in the $\theta$-term is non-zero, the expectation value of this operator should be zero. However, for a finite system of size $L^3$, the charge associated with the monopole can only be dragged to the system edge, and we therefore expect the operator to decay exponentially in the system-size for $\nu\neq 0$. When we now move on to numerical calculations, this feature will indeed be shown to hold. 

\section{Numerical simulations}
\label{sec:Numerics}
In this section, we perform a few numerical checks of our proposed order-parameters for $2+1D$ and $3+1D$ by using two simple representative lattice models.
\subsection{The Chern insulator}
 We begin with the the non-interacting Chern-insulator lattice model from Ref.~\onlinecite{Qi2006}. Specifically, we consider a square lattice with sites labeled by $\mathbf{x}$ (the lattice constant is for simplicity set to unity) and two electronic states at each site, created by the operators $c_{1\mathbf{x}}^{\dagger}$ and $c_{2\mathbf{x}}^{\dagger}$
respectively. The Hamiltonian of this model reads
\begin{align}
H&=\sum_{\mathbf{x}}\left[\mathbf{c}_{\mathbf{x}}^{\dagger}\frac{\sigma_{z}-i\sigma_{x}}{2}\mathbf{c}_{\mathbf{x}+\hat{x}}+\mathbf{c}_{\mathbf{x}}^{\dagger}\frac{\sigma_{z}-i\sigma_{y}}{2}\mathbf{c}_{\mathbf{x}+\hat{y}}+h.c.\right]\notag\\
&+2M\sum_{\mathbf{x}}\mathbf{c}_{\mathbf{x}}^{\dagger}\sigma_{z}\mathbf{c}_{\mathbf{x}},
\label{eq:ham}
\end{align}
where $\mathbf{c}_{\mathbf{x}}^{\dagger}=(c_{1\mathbf{x}}^{\dagger}\,,\,c_{2\mathbf{x}}^{\dagger})$,
$M$ is a variable parameter, and $\sigma_{x}$, $\sigma_{y}$, $\sigma_{z}$ are Pauli matrices in sub-lattice space. It can be shown analytically~\cite{Qi2006} that the ground-state of this Hamiltonian has the Hall conductance
\begin{equation}
\sigma_{H}=\frac{e^{2}}{h}\begin{cases}
\text{Sign}(M) & \text{if }|M|<1,\\
0 & \text{otherwise},
\end{cases}
\end{equation}
where the natural constants, $e$ and $h$, are re-introduced. 

To verify the asymptotic behavior of the correlation functions, Eq.~\eqref{eq:assymbehaviour}, we define the lattice operators
\begin{equation}
\label{eq:numphi}
	\phi_0(\mathbf x)
  =\eta(\mathbf{x})=\exp\left(i\sum_{\mathbf{x}^\prime\neq\mathbf{x}}\mathbf{c}_{\mathbf{x}^\prime}^{\dagger}
    \mathbf{c}^{\ }_{\mathbf{x}^\prime}\arg(\mathbf x-\mathbf x^\prime)\right),
\end{equation}
and
\begin{equation}
\phi_1(\mathbf x)=\eta(\mathbf{x})\sum_{i=1,2} c_{i\mathbf{x}}^{\dagger}, 
\end{equation}
as well as the discretized argument function
\begin{equation}
\label{eq:discArg}
	\arg(\mathbf x-\mathbf x^\prime)=\begin{cases}
				\frac{x-x^\prime+i(y-y^\prime)}{|\mathbf x-\mathbf x^\prime|} & \mathbf x\neq \mathbf x^\prime\\
				1 &  \mathbf x= \mathbf x^\prime\ .
				\end{cases}
\end{equation}

\begin{figure}[t]
\includegraphics[width=0.95\columnwidth] {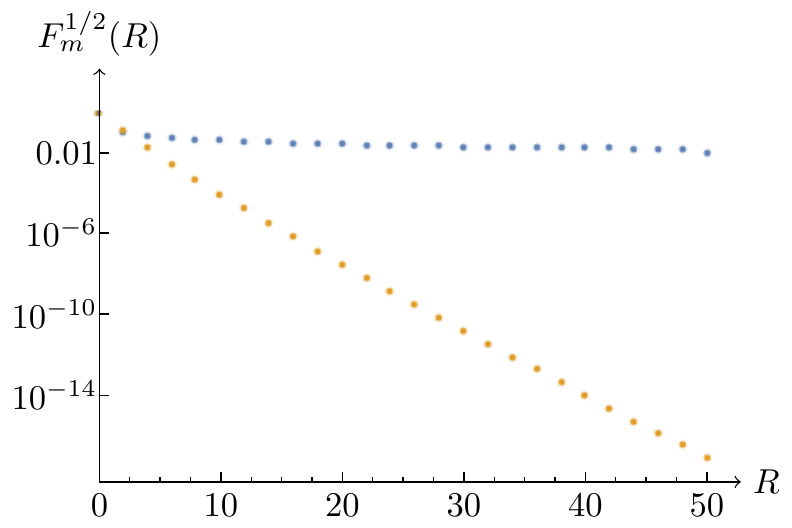}
\caption{A log-linear plot of the Chern insulator correlation functions  $F^{M=1/2}_0(R)$ (yellow dots) and $F^{M=1/2}_1(R)$ (blue dots) as functions of distance $R$ (in units of lattice sites). The mass parameter $M=1/2$ so that the Chern insulator is in a state with $\n=1$. For the exact definition of $F^{M}_m(R)$, see Eq.~\eqref{eq:NumCorr}.}
\label{fig:AlgandExp}
\end{figure}

With these definitions, we next compute the correlation functions
\begin{align}
\label{eq:NumCorr}
F^M_m(R)\equiv F^M_m(|\mathbf x-\mathbf x^\prime|)=|\Braket{\Psi_M|\phi_m(\mathbf x)\phi_m(\mathbf x^\prime)|\Psi_M}|,
\end{align}
where $\ket{\Psi_M}$ is the ground-state of the Hamiltonian~\eqref{eq:ham} for a given $M$. The functions $F^M_0(R)$ and $F^M_1(R)$ are shown in Fig.~\ref{fig:AlgandExp} for a fixed $M=1/2$ (so that $\nu=1$) on a lattice with $160\times100$ sites. The lattice size has been chosen large enough to make sure that $\mathbf x$ and $\mathbf x^\prime$ is far enough from the edge to avoid distortions from the gapless edge mode. These calculations are in accordance with our analytical results: the correlator $F^{1/2}_1(R)$ decays algebraically with $R$ while  $F^{1/2}_0(R)$ shows exponential decay in $R$. 

\begin{figure}[t]
\includegraphics[width=0.99\columnwidth] {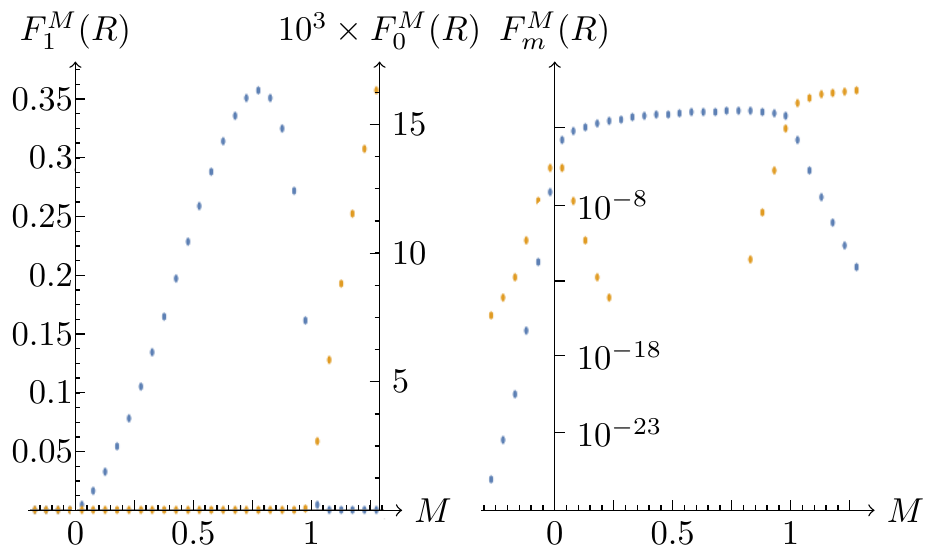}
\caption{Correlators $F^M_{0}(R)$ (yellow dots)  and $F^M_{1}(R)$ (blue dots), for fixed $R=50$ as functions of the mass parameter $M$. The left plot is linear while the right plot is log-linear. In the log-linear plot, values of $F^M_{1}(R)$ smaller than the machine precision are omitted as they are of the same order or smaller than the numerical error.
}
\label{fig:non-local-op} 
\end{figure}

We proceed by computing, in the same system, for a fixed $R=50$, $F^{M}_0(50)$ and $F^{M}_1(50)$ as functions of $M$. These results are depicted in Fig.~\ref{fig:non-local-op}. We see that $F^{M}_0(50)$  only has support in the trivial regime ($|M|>1$) while $F^{M}_1(50)$ is finite only in the the regime where $\nu=1$, ($0<M<1$). 

From these simple numerical checks, we conclude that our order parameters work as expected.

\subsection{3+1D chiral topological insulators}
As a representative state with non-zero $\theta$-term, we consider the  
$(3+1)D$ class AIII topological insulator. We use the following four-band model on a cubic lattice
\begin{multline}
\label{eq:AIIIHAm}
H=\frac{1}{2}\sum_{\mathbf{x},\mu={1,2,3}}\mathbf{c}_{\mathbf{x}}^{\dagger}(-\tau_{1}-i\sigma_{\mu}\tau_{2})\mathbf{c}_{\mathbf{x}+x^{\mu}}+h.c.\\
+ \frac{{\color{blue}2}(\mu+3)}{\color{blue}3}\sum_{\mathbf{x}}\mathbf{c}_{\mathbf{x}}^{\dagger}\tau_{1}\mathbf{c}_{\mathbf{x}}
\end{multline}
where $\tau$ and $\sigma$ are Pauli matrices in lattice and spin space respectively and $\{x^\mu\}_{\mu={1,2,3}}$ denotes the three primitive vectors of the lattice.  

\begin{figure}[ht]
\begin{center}
 \includegraphics[width=0.4\textwidth]{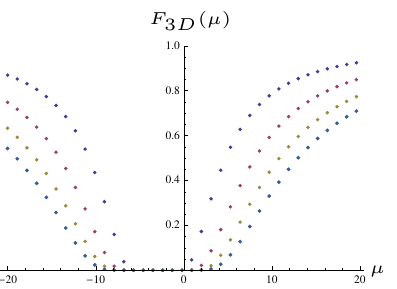}
\end{center}
 \caption{
\label{3dplot}
The ground state expectation value~\eqref{3dnum}
for various system sizes $V=12^3, 10^3,8^3, 6^3$ (top to bottom) as a function
of $\mu$.
Up to finite size corrections that vanishes with increasing
system size, the correlator  vanish in the parameter range where the $\theta$-term is non-zero, the interval $-6<\mu<0$.}
\end{figure}

The Hamiltonian~\eqref{eq:AIIIHAm} is characterized by values for $\nu$ in the $\theta$-term~\eqref{thetaterm} according to
\begin{align}
\nu=\begin{cases}
0 & 3<|\mu+3|\\
-1 & 1<|\mu+3|<3\\
2 & -4 < \mu < -2.
\end{cases}
\end{align}
To check the expected behaviour of the trivial state order parameter construction in Sec.~\ref{generalization}, we compute the correlation function 
\begin{align}
  F_{3D}(M)\equiv\lim_{R\rightarrow \infty}\left|\Braket{\Psi_\mu|\phi_0^\dagger(\mathbf x^{\prime\prime}_{\perp})\phi_0(\mathbf x^{\prime\prime}_{\perp})|\Psi_\mu}\right| \ .
  \label{3dnum}
\end{align}
We take $\phi_0(\mathbf{x})$ and the argument function as defined as in Eq.~\eqref{eq:numphi} and Eq.~\eqref{eq:discArg} repsectively. Further, $\ket{\Psi_\mu}$ is the ground state for a given value of the mass parameter $\mu$. We depict in Fig.~\ref{3dplot} this correlator as a function of $\mu$ for a few different system sizes. We see that, up to a finite size contribution that vanishes with increasing system size, $F_{3D}(\mu)$ vanishes inside the parameter range where the $\theta$-term is non-zero, the interval $-6<\mu<0$. Our computation therefore in accordance with the anticipated behaviour in Sec.~\ref{generalization}.

The operator $\eta$ is an order-parameter for the state without a $\theta$-term. With a $\theta$-term present $\eta\sim\exp(-L/\lambda)$ for some $\lambda$. However, without the chiral symmetry $\lambda\rightarrow\infty$ even without a gap closing; it can be arbitrarily large independent of the gap of the Hamiltonian. We verified that when we increase system size, the correlation function \eqref{3dnum}, in the region $-6 < \mu < 0$, typically approaches zero much slower if we break chiral symmetry.

\section{Summary and outlook}
\label{sec:Conclusions}
In this paper, our main achievement was the construction of a set of non-local order parameters for the IQH or Chern insulator states, i.e. any state characterized by its quantized value of the Hall conductance. We showed that for each such state, there exists an associated operator that can be interpreted as the insertion of local charges and $2+1D$ monopoles at two points separated in space. Such an operator was further shown to decay algebraically in the monopole separation distance only for its associated state but exponentially in all the other states. As such, their behavior characterize the insulator states, and the operators therefore constitute general non-local order parameters for these states. This generality is supported by the important fact that no microscopic input was used in our construction, but only the anomalous $2+1D$ $U(1)$ response, which can be viewed as a complementary definition of the Chern insulator.

We argued further that our construction could be generalized to other topological systems with anomalous response. We demonstrated how this could be done for the $3+1D$ topological insulators~\cite{Hosur2010,Ryu2010,Ryu2012}, with a chiral (or axial) anomalous response for the bulk and a half level Chern-Simons response for the boundary. 

Apart from providing a complementary view on topological states of matter, i.e. one in terms of a singular $U(1)$ monopole response, our construction presents a strong numerical advantage compared to computing topological invariants. Such computations are numerically expensive, especially in the presence of disorder or interactions. Our construction could therefore be very beneficial in numerical studies of strongly correlated and disordered topological systems.

We end with an outlook towards future studies. One interesting direction is to study the critical behaviour of the order parameters in the vicinity of the Hall transitions, i.e. when $\nu$ changes. Our preliminary results indicate that the operators seem to display universal behaviour in these regions. 

We anticipate that our construction can be generalized further, and in particular to $1D$ systems. It would also be interesting to study the relation between our construction and non-local order-parameters in systems beyond gapped topological systems, such as Ref.~\onlinecite{Shindou2006}. The connection to other types of non-local order parameters such as those in Ref.~\onlinecite{Matsuura2010} would also be worthwhile to explore. 

\section*{Acknowledgments}
We are grateful to T.H. Hansson for numerous enlightening discussions and valuable input on the manuscript. We also acknowledge E. Ardonne, M. Stone, E. Fradkin, and Y. Kikuchi for stimulating discussions. Parts of the results presented in this paper were obtained in collaboration with Xueda Wen who is also greatly acknowledged.

T.K.K acknowledges support by The Wenner-Gren foundations and, in part during 2018, by Stiftelsen Olle Engkvist Byggm\"{a}stare. S.R. acknowledges support by the National Science Foundation under award number DMR-1455296, and by a Simons Investigator Grant from the Simons Foundation. Parts of this work were supported by an INSPIRE Grant at University of Illinois at Urbana Champaign and Stockholm University.

\bibliographystyle{apsrev4-1-titles}
\bibliography{References}
\end{document}